\documentclass[a4paper, amsfonts, amssymb, amsmath, reprint, showkeys, nofootinbib, twoside]{revtex4-1}
\usepackage[english]{babel}
\usepackage[utf8]{inputenc}
\usepackage{svg}
\usepackage{xcolor}
\usepackage{comment}
\usepackage{gensymb}
\usepackage{float}

\usepackage{dcolumn}
\usepackage[T1]{fontenc}
\usepackage{amssymb, amsmath,amsfonts}
\usepackage{bm}% bold math
\usepackage{hyperref}
\usepackage{comment}

\definecolor{Gr}{rgb}{0,0.3,0}

\newcommand*\chem[1]{\ensuremath{\mathrm{#1}}}

\begin{document}

\title{Multigap superconductivity in Ising superconductors: The case of \chem{(LaSe)_{1.14}(NbSe_2)_m} misfit layer compounds}
   
\author{Alexandra Palacio-Morales$^{1,2,\ast}$, Tomas Samuely$^{3,4,\ast}$, Ludovica Zullo$^{5,\ast}$, Rapha\"el T. Leriche$^{1}$, Pavol Szab\'o$^{3}$, Shunsuke Sasaki$^{6}$, Cesare Tresca$^{7}$, Hugo Le Du$^{1}$, Christophe Brun$^{1}$, Fran\c{c}ois Debontridder$^{1}$, Giovanni Marini$^{8}$, Marek Kuzmiak$^{3}$, Jozef Ka\v{c}mar\v{c}\'ik$^{3}$, Laurent Cario$^{6}$, Matteo Calandra$^{8,\dagger}$, Tristan Cren$^{1,\dagger}$ and Peter Samuely$^{3,\dagger}$}

\affiliation{\vspace{0.3cm}\centering
    \begin{tabular}{c} \\ $^1$Institut des NanoSciences de Paris, Sorbonne Universit\'e and CNRS-UMR 7588, 75005 Paris, France \\ $^2$Universit\'e Paris-Saclay, CNRS, Centre de Nanosciences et de Nanotechnologies, 91120, Palaiseau, France \\ $^3$Centre of Low Temperature Physics, Institute of Experimental Physics, Slovak Academy of Sciences,\\ SK-04001 Ko\v sice, Slovakia \\ $^4$Centre of Low Temperature Physics, Faculty of Science, P. J. Safarik University, SK-04001 Ko\v sice, Slovakia \\ $^5$Institut für Theoretische Physik und Astrophysik and Würzburg-Dresden Cluster of Excellence ctd.qmat, \\Universität Würzburg, 97074 Würzburg, Germany \\ $^6$Institut des Mat\'eriaux Jean Rouxel, Universit\'e de Nantes and CNRS-UMR 6502, 44322 Nantes, France \\ $^7$CNR-SPIN c/o Diparitimento di Scienze Fisiche e Chimiche, Università degli Studi dell’Aquila, \\ Via Vetoio 10, I-67100, L’Aquila, Italy \\ $^8$Department of Physics, University of Trento, Via Sommarive 14, 38123 Povo, Italy \\
    \noalign{\smallskip}
        \small $^\ast$ These authors contributed equally to this work \\
        \small $^\dagger$ Corresponding authors: m.calandrabuonaura@unitn.it, tristan.cren@upmc.fr, samuely@saske.sk
    \end{tabular}
}

\begin{abstract}

Strong spin-orbit coupling and broken inversion symmetry in transition metal dichalcogenides give rise to Ising superconductivity, a spin-protected pairing state first identified in monolayer NbSe$_2$ through in-plane critical fields far exceeding the Pauli limit. More recently, Ising superconductivity has been proposed as a potential route to unconventional and even topological superconductivity in bulk misfit compounds. Here, we investigate the superconducting order parameter of layered misfit compounds composed of alternating transition metal dichalcogenide and rocksalt layers, which host extremely doped, electronically decoupled NbSe$_2$ sheets within a three-dimensional crystal. Using directional scanning tunneling spectroscopy on the misfit superconductors \chem{(LaSe)_{1.14}(NbSe_2)} and \chem{(LaSe)_{1.14}(NbSe_2)_2}, we uncover a strongly anisotropic multigap superconducting state: a fragile gap on the $\Gamma$-centered Fermi-surface pocket coexists with a robust, intrinsic gap on the K and K$'$ pockets. These features are in quantitative agreement with momentum-resolved gaps $\Delta(\mathbf{k})$ obtained from anisotropic Migdal-Eliashberg calculations. The marked fragility of the $\Gamma$-centered gap, combined with the strong sensitivity of the critical temperature to non-magnetic disorder, points to pairing beyond conventional $s$-wave symmetry, potentially involving a topological order parameter. These results establish NbSe$_2$-based misfit compounds as a tunable bulk platform for multigap, unconventional superconductivity, with Ising protection offering a promising route toward topological pairing.

\end{abstract}

\maketitle

\section{Introduction}

The interplay of reduced dimensionality, strong spin–orbit coupling (SOC) and broken inversion symmetry in layered materials has opened new perspectives for unconventional and topological superconductivity. A prominent example is Ising superconductivity, first identified in monolayers of transition metal dichalcogenides (TMDs) such as NbSe$_2$ and gated MoS$_2$ \cite{xi_ising_2016, lu_evidence_2015, xi_ising_2015}. There, strong SOC locks electron spins out-of-plane near the K and K$'$ valleys, protecting Cooper pairs against in-plane magnetic fields and allowing upper critical fields to exceed the Pauli limit by more than an order of magnitude \cite{xi_ising_2016, saito_ising_2016}. This spin–valley–momentum structure has also been proposed to favor unconventional and even topological pairing channels, including mixed singlet–triplet amplitudes and equal-spin pairing \cite{qian_solid_2014, fei_edge_2017, ugeda_collective_2022}.

Theoretical studies of monolayer NbSe$_2$ predict strongly momentum-dependent pairing, with dominant gaps on the K and K$'$ Fermi surface pockets and a much smaller gap on the $\Gamma$ pocket induced by interband proximity \cite{das_anisotropic_2023, wicramaratne_prx_2020}. Experimentally, signatures consistent with competing triplet channels and collective modes beyond conventional $s$-wave behaviour have been reported in few-layer NbSe$_2$ \cite{kuzmanovic_spintriplet_2022, ugeda_collective_2022}, yet the rapid suppression of Ising protection with increasing thickness in 2H stacking has limited exploration of this multigap, potentially unconventional pairing in bulk crystals.

Layered misfit compounds composed of alternating TMD and rocksalt layers offer a distinct route to stabilizing two-dimensional, SOC-protected superconductivity in a bulk environment \cite{samuely_misfit_2021}. In these materials, extreme charge transfer and reduced interlayer coupling decouple the NbSe$_2$ sheets, so that the resulting crystal behaves as a stack of highly doped, effectively isolated TMD layers. We have shown that $(\mathrm{LaSe})_{1.14}(\mathrm{NbSe}_2)$ and $(\mathrm{LaSe})_{1.14}(\mathrm{NbSe}_2)_2$ exhibit an in-plane upper critical field far exceeding the Pauli limit, and the microscopic mechanism that preserves spin-momentum locking deep into the bulk has since been deciphered \cite{samuely_misfit_2023}. These properties make misfit compounds a natural bulk platform to explore the momentum-dependent, multigap superconductivity predicted in monolayer NbSe$_2$, and to test whether it survives, and evolves, in a three-dimensional crystal.

Here, we investigate the superconducting energy gap in these misfit superconductors. Using directional scanning tunneling spectroscopy, we uncover a strongly anisotropic multigap superconducting state, with a fragile gap on the $\Gamma$ pocket and a robust intrinsic gap on the K and K$'$ pockets. Comparison with anisotropic Migdal-Eliashberg calculations shows quantitative agreement with a momentum-resolved gap $\Delta(\mathbf{k})$. The pronounced fragility of the $\Gamma$ pocket gap and its sensitivity to non-magnetic disorder point to pairing beyond conventional $s$-wave symmetry. Together with the valley-locked spin structure inherited from Ising protection, these results establish NbSe$_2$-based misfit compounds as a tunable bulk platform for multigap, unconventional superconductivity, with a possible topological character.

\section{Results}
\subsection{Crystal and electronic band structure}

The misfit materials consist of transition metal dichalcogenide layers sandwiched with rock salt chalcogenide layers \cite{rouxel_chalcogenide_1995}, namely \chem{(MX)_{1+x}(TX_2)_n} (M=Pb, Sn, Sb, Bi or Rare-Earth, X=Chalcogene, T=Transition metal and n = 1, 2) which exhibit an intergrowth structure formed by the alternated stacking of MX planes (having NaCl structure) with TMD layers (\chem{CdI_2} or \chem{NbS_2} structure). The misfit materials \chem{(LaSe)_{1,14}(NbSe_2)_{m=1,2}} \cite{Roesky_structure_1993} consist of a regular alternation of a quasi-hexagonal (H) NbSe$_2$ monolayer (m=1) or NbSe$_2$ bilayer (m=2) with a trigonal prismatic structure and a rocksalt two-atom thick LaSe layer (Fig. 1a-b). Each NbSe$_2$ and LaSe sublattice has its own set of cell parameters noted as a$_v$, b$_v$ and c$_v$ with $v=1$ and 2, respectively. The lattice of NbSe$_2$ layers is not perfectly hexagonal and is slightly compressed along the $\vec{a}_1$ direction; the NbSe$_2$ sublattice can be described by a centered orthorhombic cell with in-plane lattice vectors a$_1 = 3.437$~\AA~and b$_1 \approx 6$~\AA. The LaSe sublattice has similar in-plane lattice parameters a$_2\approx$ b$_2\approx 6$~\AA, so it is denoted as tetragonal (T). In the following, we refer to the compounds as 1T1H and 1T2H, respectively. Both NbSe$_2$ and LaSe layers share the same $\vec{b}$ lattice vector and are, thus, commensurate in this direction. The vectors $\vec{a}_1$ and $\vec{a}_2$ share the same direction, but the ratio of their norm is irrational, making them incommensurate in the $\vec{a}$ direction; however, they can be considered as almost periodic in the $\vec{a}_2$ direction with an approximate commensurate lattice vector $\vec{m}$ (m = 7a$_1\approx$ 4a$_2$), see Fig. 1c. 

 \begin{figure*}[ht]
        \includegraphics[scale=0.35]{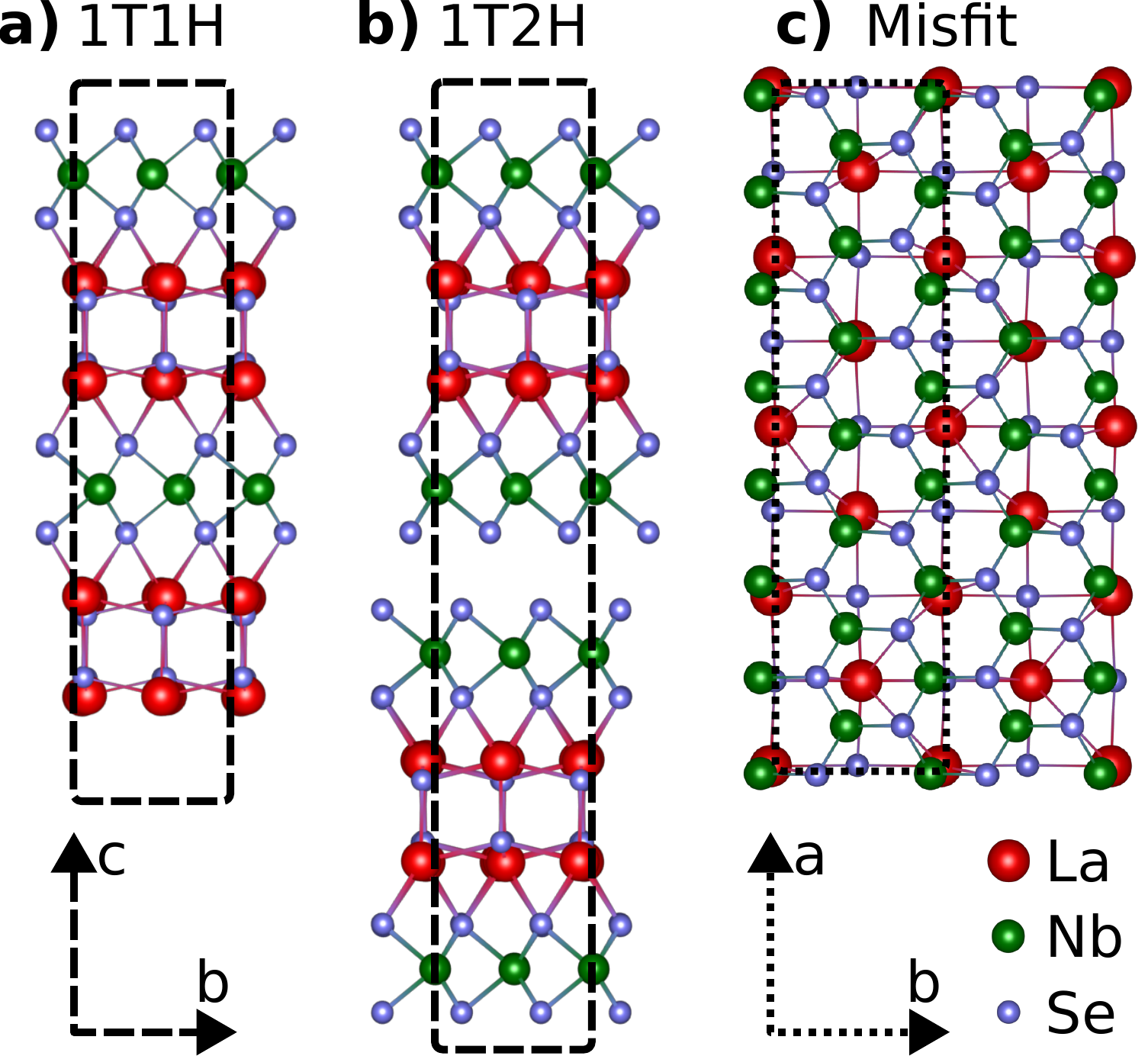} 
        \caption{
        \textbf{a)} Lateral view of the stacking of the \chem{(LaSe)_{1,14}(NbSe_2)} compound.
        \textbf{b)} Lateral view of the stacking of the \chem{(LaSe)_{1,14}(NbSe_2)_2} compound; van der Waals gap appears in between the \chem{NbSe_2} layers.
        \textbf{c)} Top view of both misfit compounds; the approximate commensurate supercell is shown by a black dotted line.
        }
        \label{Figure1}
    \end{figure*}

Strong iono-covalent bonds form at the contact between \chem{NbSe_2} and LaSe due to a large charge transfer between LaSe and \chem{NbSe_2} bilayers. It was shown by Hall measurements, ARPES, and DFT calculations \cite{samuely_misfit_2023} that in 1T2H the LaSe layer donates about 0.6~electron per \chem{NbSe_2} chemical unit corresponding to the Fermi level energy shift of 0.3~eV towards higher energy of the Nb-derived pockets in comparison with monolayer \chem{NbSe_2}. 1T1H is even more heavily doped, and the Fermi level is shifted by 0.43~eV still not completely filling the hole Nb band. Both systems share a similar electronic band structure with \chem{NbSe_2} monolayer wich indicates a rigid band shift at play (Fig.2a). Moreover, ARPES proved that the misfits are quasi-2D systems with a very little dispersion of the electronic band along the z-direction \cite{samuely_misfit_2023}. Being the electronic equivalent with \chem{NbSe_2} monolayer with Ising spin-orbit splittings around the K(K’) points in the Brillouin zone makes the misfits Ising superconductors. As the chemical potential comes close to the top of the pockets at K and K’ where the Ising splitting is maximum, we expect that this would result in a drastic enhancement of the Ising spin-orbit coupling effect on the superconducting pairing.

 \begin{figure*}[ht]
        \includegraphics[scale=.35]{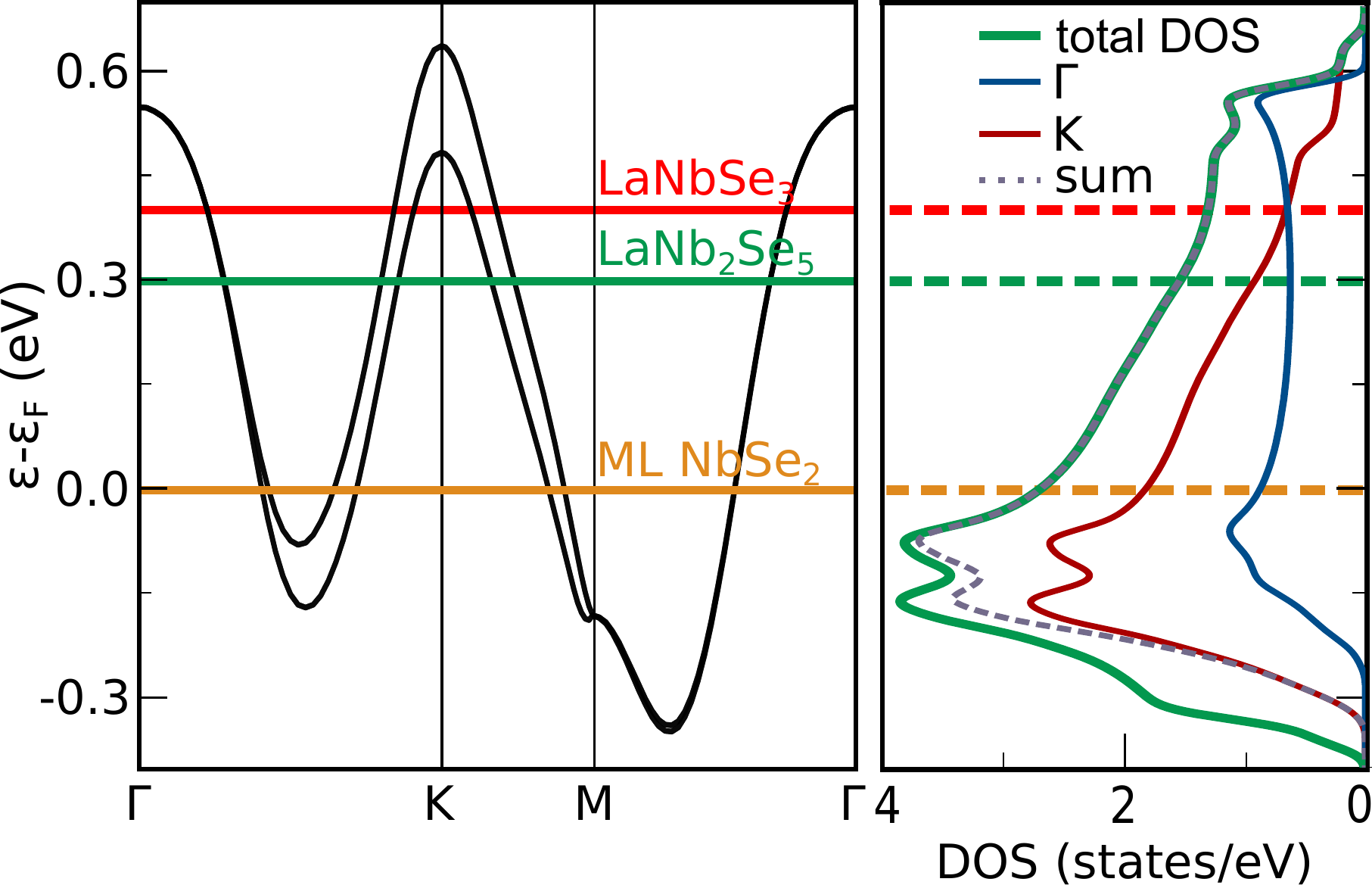} 
        \caption{
        Left: band structure calculation in the rigid band approximation showing the chemical potential in monolayer \chem{NbSe_2} (yellow), in \chem{(LaSe)_{1,14}(NbSe_2)_2} (green), and in \chem{(LaSe)_{1,14}(NbSe_2)} (red). Right: Total density of state (green) and partial DOS at $\Gamma$ (blue) and K (red).
        }
        \label{Figure2}
    \end{figure*}

In the 1T2H compound \chem{(LaSe)_{1,14}(NbSe_2)_2}, adjacent \chem{NbSe_2} layers are van der Waals bound; therefore, the material cleaves in between \chem{NbSe_2} layers. The 1T1H compound, at variance with the 1T2H compound, is not a van der Waals stacking; all the interlayer bonds are iono-covalent. In the most cases cleavage leads to a monolayer (ML) \chem{NbSe_2} terminated ($ab$)-plane surface after scanning removes some residues of LaSe, however, we have occasionally observed an unstable tetragonal LaSe termination \cite{samuely_misfit_2023}. A typical STM topography map of 1T1H surface of an in-situ cleaved sample is presented in Fig.3a. Since the surface layer is 1H-NbSe2, the atomically resolved STM images show the Se atoms of \chem{NbSe_2}. The same surface appears after cleaving the 1T2H samples. 

Importantly, in the case of 1T1H we were able to prepare well resolved surfaces perpendicular to the ($ab$)-plane by breaking the crystal. The fractured side surface displays a stripe-like topography characteristic of the layered crystal structure (Fig. 3b). The stripe heights vary, while the minimum stripe width is approximately equal to the thickness of a single structural layer. Fig. 3c shows a height profile extracted along the line indicated in the topographic image of the fractured side surface. 

\subsection{Superconducting density of states}
The 1T2H and 1T1H samples we study here have nominal $T_c \approx 4.1$~K (Fig.4a) and 1.2~K (Fig.4d), respectively.  While in the case of 1T1H it is in agreement with our previous studies \cite{samuely_misfit_2021}, in the case of 1T2H it is less than previously reported $T_c \approx 5.5$~K. In 1T2H we indeed observe strong variations of the critical temperature from sample to sample, while they exhibit nearly the same nominal composition according to our fluorescent X-ray spectroscopy measurements. The onset $T_c$ varies from 5.3~K down to 1.5~K depending on the batch. We observe a clear correlation between the residual-resistance ratio (RRR) and the critical temperature, indicating that the amount of disorder in the compounds correlates with the reduction of $T_c$ (Fig.4a, lower inset). This is not expected in a conventional BCS superconductor, which is immune to weak disorder according to Anderson’s theorem \cite{anderson_1959}. This kind of correlation of RRR with $T_c$ is usually observed in non-conventional superconductors such as heavy fermions, which exhibit sign-changing order parameters that are fragile upon disorder \cite{Hirschfeld_2011}. Non-magnetic scatterers in a sign-changing order parameter system produce similar effects as magnetic defects in a singlet s-wave superconductor; a high level of scatterers will ultimately reduce $T_c$ to zero like magnetic impurities do in a conventional superconductor \cite{abrikosov_1961}.

 \begin{figure*}[ht]
        \includegraphics[scale=0.35]{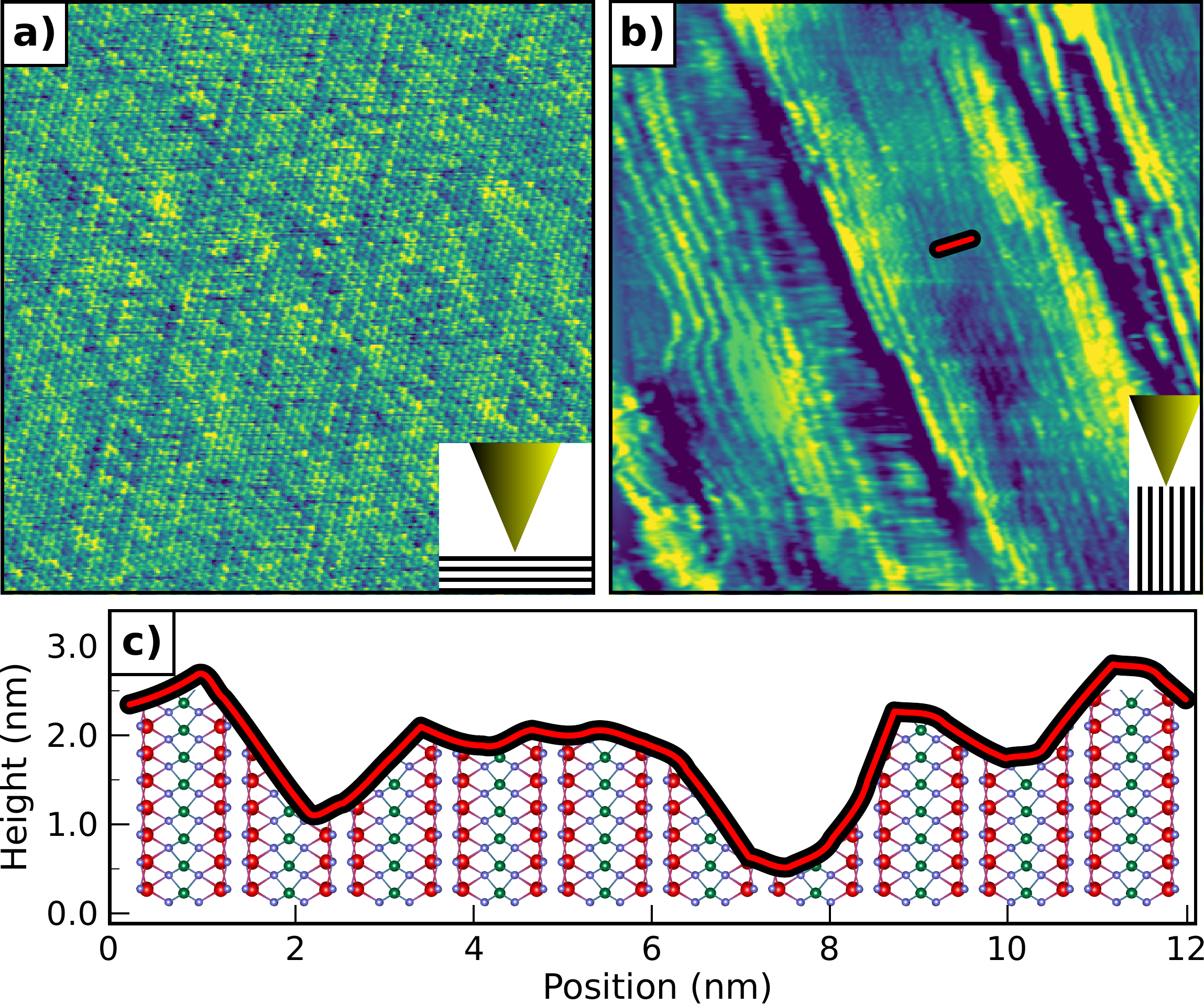} 
        \caption{
        \textbf{a)} STM topography of the surface of the \chem{(LaSe)_{1,14}(NbSe_2)} compound after scanning removed the LaSe top layer. (20 $\times$ 20 nm$^2$)
        \textbf{b)} STM topography of the fractured side surface of the \chem{(LaSe)_{1,14}(NbSe_2)} compound. (200 $\times$ 200 nm$^2$)
        \textbf{c)} Topography profile along the red line shown in (b). 
        }
        \label{Figure3}
    \end{figure*}

To gain more insight on the nature of superconducting pairing in the material, we performed scanning tunneling spectroscopy experiments at 300~mK. Typical spectra measured at various positions on the atomically resolved surfaces as shown in Fig.3a are displayed in Fig.4b panel for 1T2H and in Fig.4e for 1T1H samples by red lines. In addition, the BCS spectra expected for a superconductor with a critical temperature of respectively 4~K and 1.3~K are shown (gray curves). There is a clear disagreement between the size of the measured gap and the expected one. In the case of 1T2H the observed gap with the half-peak-to-peak distance is about $\Delta_p = 0.25-0.3$~meV while the BCS gap should amount approximately to 0.62~meV. Moreover, the gap should be totally open at 300~mK, even considering the reduced gap. Instead, we observe many in-gap states with a gap filling varying from 50\% to 80\%. Such gap filling is usually not expected in a BCS superconductor except when magnetic impurities are present \cite{abrikosov_1961, steglich_1987}. Remarkably, a very similar situation holds for the 1T1H sample. As shown in Fig.4e only a small gap-like structure is opening at any point of the atomically resolved surface with the half of the peak-to-peak distance equal to about 0.2~meV while the BCS gap should amount to about 0.23~meV and be almost fully opened around zero-bias.

 \begin{figure*}[ht]
        \includegraphics[scale=0.35]{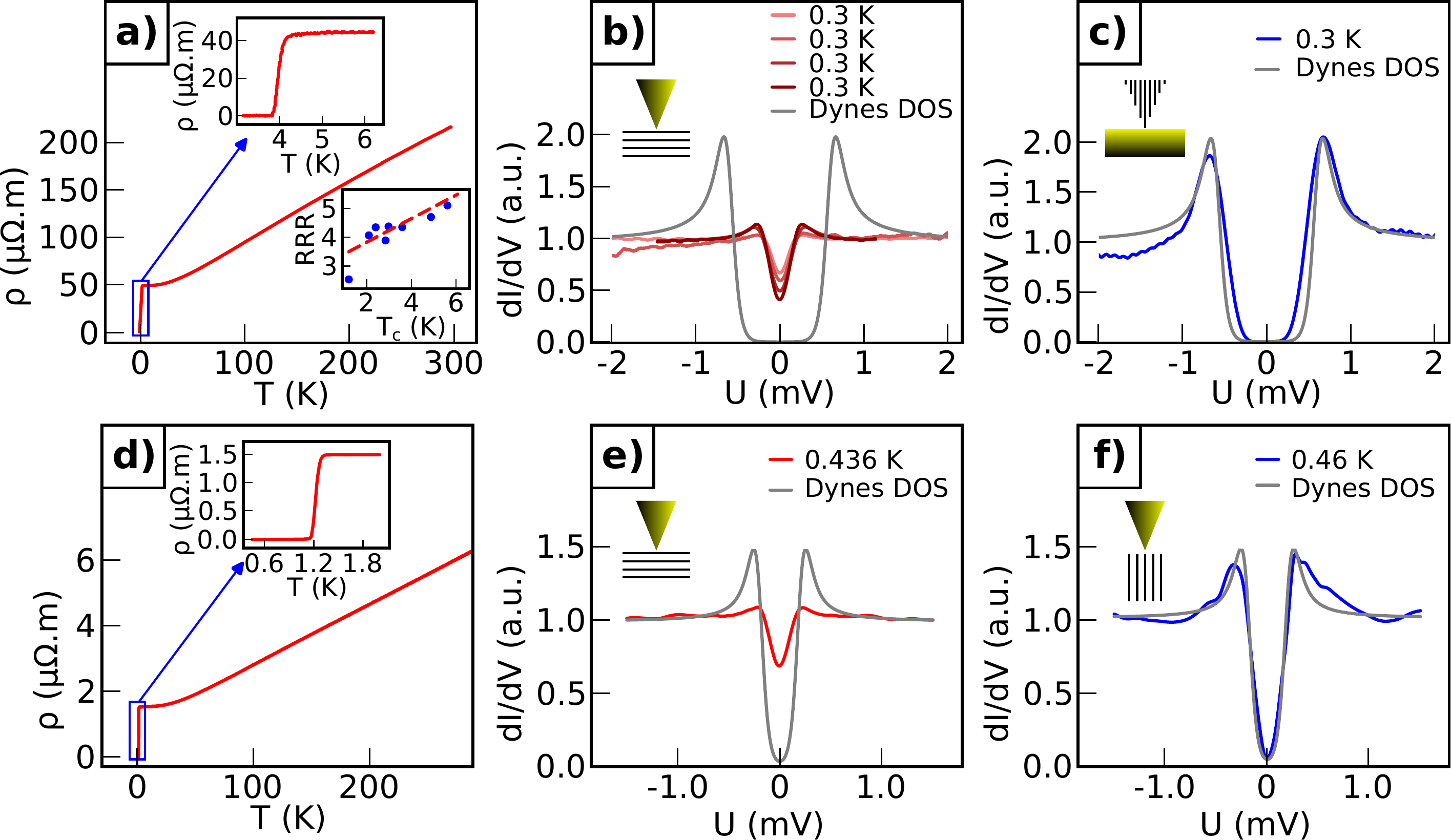} 
        \caption{
        \textbf{a)} Resistivity as function of temperature for the 1T2H compound. A zoom close to the superconducting transition is shown in the upper inset. The lower inset shows the correlation of the RRR with the critical temperature. 
        \textbf{b)} The pink-red curves show a typical set of differential conductance spectra acquired at different locations in c-axis tunneling (geometry shown in the inset). The gray curve is the expected BCS gap according to the $T_c$. 
        \textbf{c)} In blue, conductance spectrum measured on the fractured side surface by gluing an 1T2H sample on the tip (see inset). In Gray, the expected BCS gap is shown.   
        \textbf{d)} Resistivity as function of temperature for the 1T1H compound. A zoom close to the superconducting transition is shown in the inset.
        \textbf{e)} The red curve shows a typical differential conductance spectrum acquired in c-axis tunneling (geometry shown in the inset). The gray curve is the expected BCS gap according to the $T_c$. 
        \textbf{f)} In blue, conductance spectrum of 1T1H measured on the fractured side surface (see inset). In Gray, the expected BCS gap is shown.
        }
        \label{Figure4}
    \end{figure*}

Intrigued by the fact that the observed gap measured in c-axis tunneling could not account for the $T_c$ we probed the tunneling in another direction. Indeed, c-axis tunneling probes mainly the states close to $\Gamma$ point and as the inner pocket Fermi surface shrinks close to $\Gamma$, it will dominate the tunneling. Conversely, the pockets close to K and K’ are also shrinking and getting more distant from $\Gamma$ in this configuration, which means that they will have a negligible contribution to the c-axis tunneling. Previous works on \chem{2H-NbSe_2} \cite{noat_quasiparticle_2015} and \chem{MgB_2} \cite{Iavarone_2002, Rodrigo_2003} have shown that probing samples in other crystalline directions, in particular along the edge, could help reveal hidden gaps in multigap superconductors. Therefore, we performed measurements on the fractured side surface of 1T1H sample. As shown in Fig.3b, remarkably, we achieved almost atomic resolution in STM topography. In this configuration, the tunneling selectivity is more favorable to the K or K' direction, since there are some states with $k_{\parallel}=0$ at E$_F$ around K and K' pockets for this orientation. As shown in the blue curve of Fig.4f, the resulting tunneling spectra reveal much larger and full gap, as expected for a BCS tunneling conductance (gray curve). 

The 1T2H sample was too thin to perform such scanning tunneling spectroscopy measurements on its edge. We adopted a method that was used for probing \chem{MgB_2} grains \cite{Giubileo} and \chem{2H-NbSe_2} \cite{noat_quasiparticle_2015}, it consists of gluing the sample on the STM tip, with the sample plane parallel to the tip axis. The tunnel junction is obtained by scanning over a flat Au(111) sample. The resulting conductance is shown on the blue curve of Fig.4c. We observe a hard gap with a peak-to-peak width that corresponds perfectly to the width of a BCS gap with $T_c$ of 4~K.

\subsection{Density functional theory results}

We now turn to discuss the theoretical insights into the critical temperature and superconducting gap of these NbSe$_2$-based misfit layer compounds based on ab initio electron-phonon coupling calculations. 

We can predict the vibrational properties and electron-phonon coupling of large misfit supercells, including those of the 1T1H and 1T2H compounds, using our theoretical model, which conceptualizes misfits as a collection of field-effect transistors.
Thanks to the difference in work function (W) between rocksalts and TMDs (W(RS)<W(TMD)), misfit heterostructures can be regarded as a field effect transistor device built by stacking heavily electron-doped TMDs \cite{Zullo}. 

Therefore, we can model the misfit by replacing the RS unit with positively charged gates within the field-effect transistor setup developed in Refs.\cite{Brumme2015,Sohier2017}. This method has been proven to successfully predict both the charge density wave (CDW) collapse of NbSe$_2$ in the 1T2H bulk misfit \cite{Zullo2} and its doping-induced CDW behaviour as a function of the mutual La and Pb concentration in the $(\text{La}_x\text{Pb}_{1-x}\text{Se})_{1.14}(\text{NbSe}_2)_2$  misfit series \cite{le_du_2026_CDW}.  

In this work, we investigate both a mono- and bi-layer NbSe$_2$ in a double-gate field effect setup to model the 1T1H and 1T2H misfit compounds. We choose the amount of charge on the gates to match the experimental Fermi surface determined by ARPES in Ref.\cite{samuely_misfit_2023}. More details of the calculations can be found in the SI.

We calculate the superconducting properties of the misfit from the electron-phonon coupling, as implemented in the EPI\textit{q} software \cite{EPIQ} (see the SI).

First, the superconducting critical temperature is obtained from the Allen-Dynes formula \cite{Allen-Dynes} by solving the Eliashberg equations. We assume a high value for the Coulomb pseudopotential ($\mu^{*}=0.15$), in agreement with previous ab initio studies on the reduced effectiveness of Coulomb renormalization arising from the electronic structure of NbSe$_2$ \cite{Sanna2022}. We estimate a critical temperature of T$_{c}$$=1$ K and $5.4$ K for the 1T1H and 1T2H compounds, respectively. These values are in good agreement with the experiments, providing a further confirmation that, not only the electronic and vibrational properties \cite{Zullo,Zullo2}, but also the electron-phonon coupling of the misfit is fairly captured by the FET modeling. This statement holds in the limit in which the RS modes are not dominant in the phonon density of states of the misfit, which is the case for the LaSe compounds, as demonstrated in Ref. \cite{Zullo2}.

\begin{figure*}[ht]
        \includegraphics[scale=0.35]{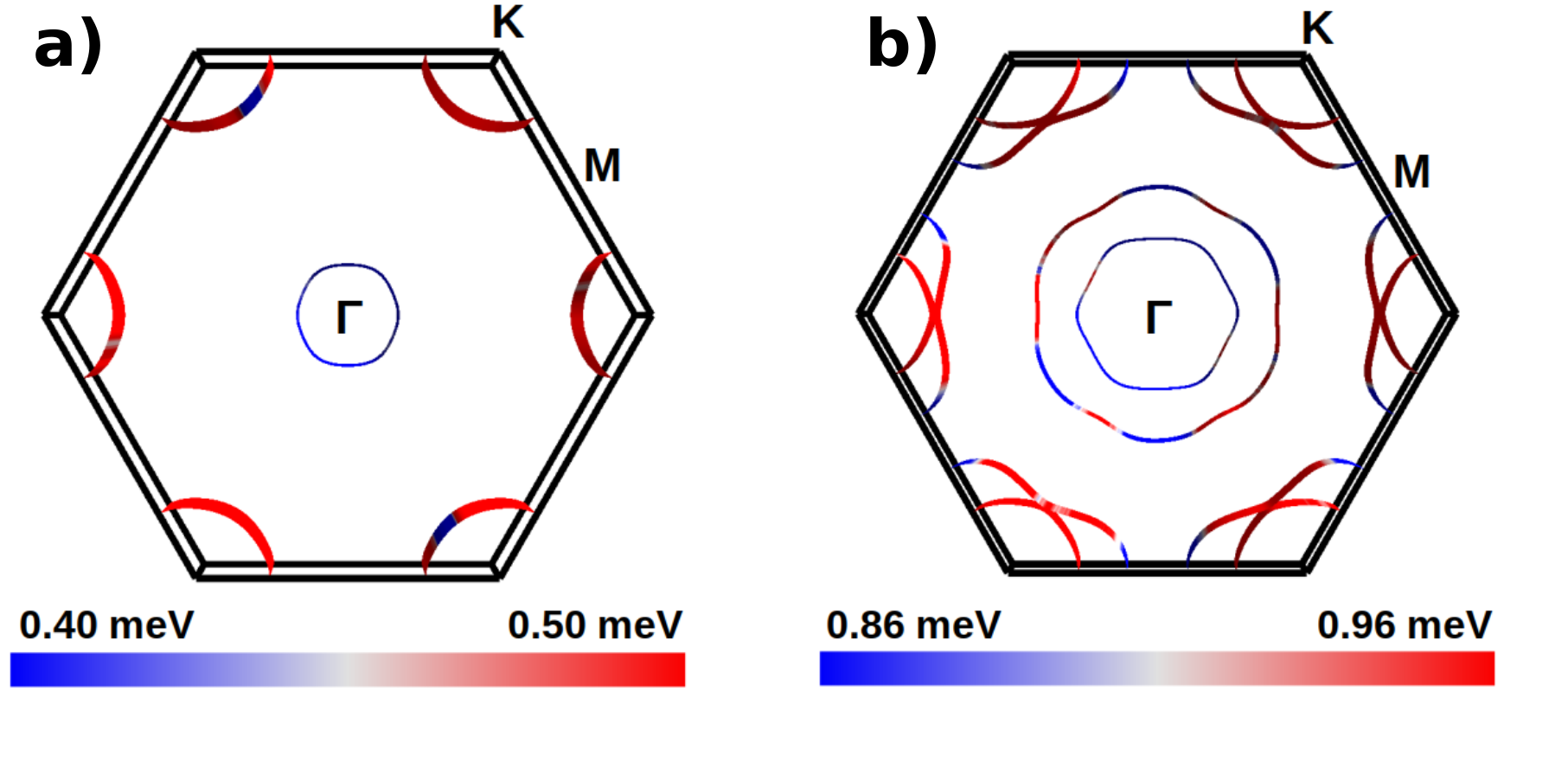} 
        \caption{
         Calculated momentum-resolved superconducting gap within the Migdal-Eliashberg theory framework for \textbf{a)} 1T1H and \textbf{b)} 1T2H compound. 
         The color bar represents  the superconducting gap centered around its mean value.  
        }
        \label{Figure5}
    \end{figure*}

To determine the gap anisotropy of the NbSe$_2$, we calculate the momentum-resolved superconducting gaps $\Delta(\mathbf{k})$. This is done by solving the anisotropic Migdal-Eliashberg equations \cite{MigdalEliashberg}. 
Fig.5 shows the results for a) the 1T1H and b) 1T2H compound.  
The plot colorbars represent the distribution of the superconducting gap over the Brillouin zone, whose value is centered on the average gap, namely $0.45$ and $0.91$ meV for the 1T1H and the 1T2H, respectively.

To understand the difference between the two misfit compounds we start from the shape of the Fermi surface.
In both cases, the Fermi surface features two types of electron pockets: one centered at $\Gamma$ and another located around the K and K$^{'}$ valleys.
It must be pointed out that in our calculations we neglect the spin-orbit coupling. However, even if the effect of the SOC is to split the degenerate bands, the density of states of the electrons at the Fermi level does not change; therefore, the electron-phonon coupling is less affected. 
Indeed, our results align with the experimental ARPES Fermi surfaces we measured previoulsy \cite{samuely_misfit_2023}; thus, our model effectively captures the key features of the misfit electronic states at the Fermi level.

Now we turn to the anisotropy of the superconducting gap distribution across the Fermi surface. As illustrated in Fig.5, the gap distribution is non-homogeneous in both cases. Specifically, the $\Gamma$ pocket exhibits a lower gap value compared to the K valleys. 
By estimating the critical temperature through the BCS formula \(T_{c}=\Delta/1.76k_{B}\) taking into account the anisotropic Migdal-Eliashberg gap value, we found critical temperatures of $3$ K for the 1T1H and $6$ K for the 1T2H compound. For the 1T2H compound, this result is in good agreement with the isotropic Allen-Dynes value, while the value for the 1T1H compound is three times larger, highlighting its pronounced anisotropic nature.
Qualitatively, the 1T1H shows a more pronounced difference between the pockets than the 1T2H. This observed difference can be attributed to the stronger localization of the $\Gamma$ pocket at the center, which is influenced by the high charge transfer in the misfit. In contrast, the doping in the 1T2H leads to a more spatially extended pocket around $\Gamma$ on the Fermi surface, resulting in greater mixing between points in the Brillouin zone.

This point raises the importance of the charge transfer in the misfit: it is a key feature not only for the electronic and vibrational properties of this class of materials, but also for the regulation of the electron-phonon coupling ($\lambda$).
Our calculations indicate that the density of states at the Fermi level N(E$_F$), which is tuned by doping in the misfit, plays a crucial role in the superconducting behaviour. The ratio of $\lambda/$N(E$_F$) remains nearly constant when transitioning from the 1T1H to the 1T2H compounds.
Consequently, the variation in electron-phonon coupling is primarily influenced by the number of available states at the Fermi level, which is greater in the bilayer due to the higher electron count.
As a result, our calculations indicate that the $T_c$ is enhanced by a larger value of lambda in the 1T2H compound with respect to the 1T1H case.

The ability to tune charge transfer in the misfit through control knobs, such as varying the rocksalt units, alloying them in the misfit, and changing the number of TMD layers, is therefore pivotal for superconductivity.

\subsection{Discussion }

The anisotropic Migdal-Eliashberg equations \cite{MigdalEliashberg} were solved assuming a singlet s-wave order parameter, however many other possible order parameters could be at play. The weakness of the gap for the states located at $\Gamma$ pocket, manifested by a small gap and a high zero bias conductance, together with the correlation of the critical temperature with the RRR (Figure 4a, Inset) might be explained by either magnetic impurities in a s-wave superconductor or by the effect of non-magnetic impurities in a non-conventional superconductor. We did not find any traces of magnetic impurities by X-rays fluorescence, but we cannot exclude the presence of a minute amount of contaminant. The alternative relies upon non-magnetic impurities, such as La vacancies, that were found in our samples. A strong effect of non magnetic impurities that manifest as a reduced $T_c$ or by in-gap states around the Fermi level could be explained quite naturally by a non-conventional order parameter. 

We can grasp some hint on the possible symmetry of the superconducting order parameter by taking into account the symmetry of the TMDs layers.
The misfit's 1H-NbSe$_2$ layers have nearly C$_{3V}$ symmetry and a sizable Rashba spin-orbit due to heavy charge transfer from the rocksalt. A wealth of order parameters is possible in such a case as shown by Haniš et al. \cite{haniš_distinguishing_2024}.  Among them, we may exclude chiral order parameters such as triplet $p_x\pm i p_y$ or singlet $d_{x^2-y^2}\pm i d_{xy}$ since they would manifest some edge states at step-edges that would fill the gap, while instead we observe fully developed gap on the sample's edges.

Among the available non-conventional order parameters that were proposed for 1H-NbSe$_2$ with both Ising and Rashba spin-orbit coupling we may consider a singlet order parameter$(\lvert\uparrow,\downarrow\rangle - \lvert\downarrow,\uparrow\rangle)/\sqrt{2}$ with nodal $i$-wave pairing ($\Delta^{s}_{A_2}$) (We use Hani\v{s}et al. terminology~\cite{haniš_distinguishing_2024}). We may also consider triplet pairing with a$(\lvert\uparrow,\downarrow\rangle + \lvert\downarrow,\uparrow\rangle)/\sqrt{2}$spin part, which is compatible with pure Ising pairing; it corresponds to a triplet $d$-vector vector along $z$ with a nodal triplet$f$-wave pairing ($\Delta^{t,z}_{A_1}$ and $\Delta^{t,z}_{A_2}$). Other order parameters can also be at play if we consider a substantial Rashba spin-orbit coupling. More experiments, involving for instance quasi-particle interferences around defects, are needed in order to resolve this order parameter symmetry issue.

\subsection{Conclusion }

In conclusion, our results demonstrate the anisotropic nature of the superconducting gap in NbSe$_2$ within misfit layer compounds. We find a small and fragile gap around the $\Gamma$ pockets while the K and K' pockets exhibit a well developed hard gap with a width in accordance with the critical temperature. Our DFT calculations predict a stronger pairing around K and K' pockets compared to $\Gamma$ pockets in accordance with our observation. Our electron-phonon coupling DFT calculations predict superconducting critical temperatures that align with experimental observations in both 1T1H and 1T2H compounds. The fragile gap around $\Gamma$ with the correlation between the $T_c$ and RRR suggest a non-chiral unconventional pairing. Our experiments could not discriminate between all the possible order parameters. Quasiparticle interference around strong non-magnetic scatterers could shed more light on the pairing symmetry, in particular, it could permit discriminating between nodal and nodeless order parameters.

\vskip 0.5cm
{\bf Author contributions.}
TS, PSz, MK, JK and PS contributed to the STM/STS measurements of 1T1H and the transport measurements on both compounds.
APM, RL, CB, TC contributed to the STM/STS measurements of 1T2H compounds, with the technical support of FD.
LZ, GM, CT and MC performed DFT calculations. SS and LC grew the crystals and analysed them by X-ray diffraction and X-Ray spectroscopy.
HLD and TC contributed to the discussion on the symmetry of the order parameter.

\vskip 0.5 cm
{\bf Competing interests.}
The authors declare no competing interests.

\bibliography{references}

\end{document}

\title{Supplementary information for Multigap superconductivity in Ising superconductors: The case of \chem{(LaSe)_{1.14}(NbSe_2)_m} misfit layer compounds}
\author{Authors}
\date{June 2026}

\maketitle

\section{Methods}  

     \paragraph{Theory} We calculate the vibrational properties of the misfit within the field-effect modeling by means of density functional perturbation theory (DPFT), in the linear response regime \cite{Baroni2001}, as implemented in the Quantum ESPRESSO (QE) code \cite{QE}. 

    We employ ultrasoft pseudopotentials from the Vanderbilt distribution for Nb, including semi-core states \cite{Vanderbilt1990}, while for Se we use norm-conserving pseudopotentials with empty d-states in valence. The kinetic energy cutoffs for the plane-wave basis set and for the charge density of NbSe$_{2}$ are set to $50$ and $500$ Ry, respectively. The Brillouin zone (BZ) integration is performed with a Monkhorst-Pack grid of 30$\times$30$\times$1 k-points and Methfessel-Paxton (m-p) smearing of $0.005$ Ry. DFPT is performed on a uniform 12$\times$12$\times$1 phonon-momentum grid.

    As detailed in Ref. \cite{Zullo}, we modeled the misfit as a collection of field effect transistors (FET) by means of the 2D materials FET-setup developed in Refs. \cite{Brumme2015,Sohier2017}. 
    Within this method, the effect of the misfit 1T1H (1T2H) structure onto the NbSe$_2$ layer is modeled by a monolayer (bilayer) TMD sandwiched between two uniformly positive charged gates.
    Each charged gate replaces the RS subunit and has a positive charge per Nb corresponding to $0.68$ ($0.6$) times the modulus of the electronic charge for the monolayer (bilayer) NbSe$_2$, being the amount of charge transferred from each RS in the misfit \cite{samuely_misfit_2023,leriche_misfit_2021}. In the case of the 1T2H we shifted the Fermi energy of $0.15$ eV to better match the experimental Fermi surface of the misfit. This shift, which corresponds to a charge transfer of $0.46$ electrons per Nb atom in each layer, does not result in a qualitative change in the vibrational properties of NbSe$_2$, as we proved in Ref. \cite{le_du_2026_CDW}. Specifically, at this doping the charge density wave of NbSe$_2$, present below $0.4$ electrons per Nb atom, has already collapsed, and no dynamical instability is present.
    
    A Coulomb long range interaction cutoff is placed at z$_{cut}$ = c$/2$ with c being the unit-cell size in the direction perpendicular to the plane: c is set opportunely to $16$ and $26$ \AA for the mono- and bi-layer, respectively \ . NbSe$_{2}$ is centred around z=$0$.
    
    To model both the 1T1H and 1T2H misfits we use a double gate setup, placing charged plates modelling two gate electrodes at the coordinate $z_{bot}=-0.254c$ ($z_{bot}=-0.266c$) and $z_{top}=0.254c$ ($z_{bot}=-0.266c$) for the monolayer (bilayer) NbSe$_2$. The charge of the gate is equal and opposite to the one of each RS layer in the misfit to ensure charge neutrality. For each system a potential barriers V of height $2.5$ Ry is placed before the gates at $z_{V}=z_{bot}+0.1$ ($z_{V}=z_{top}-0.1$) in order to confine the atoms between the gate electrodes.

\begin{figure*}[h]
        \includegraphics[scale=0.35]{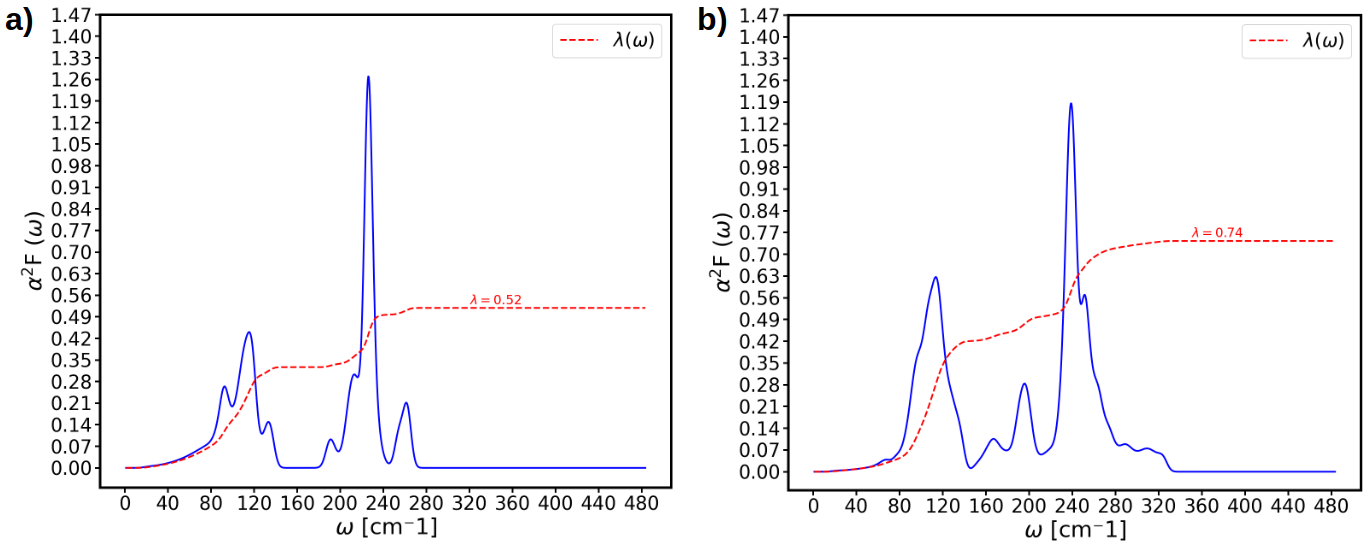} 
        \caption{
         Calculated Eliashberg function ($\alpha^{2}F$, bule solid line) for \textbf{a)} 1T1H and \textbf{b)} 1T2H compound. 
         The red dashed line is the total electron-phonon coupling ($\lambda(\omega)$).
        }
        \label{FigureSI}
    \end{figure*}

    The superconducting calculations are performed via Wannier interpolation of the electron-phonon matrix elements as implemented in the Wannier90 \cite{Wannier90} code. Within this approach, the electron-phonon matrix elements are first calculated on a coarse $12\times12\times1$ q-grid and $30\times30\times1$ k-grid, and then Wannier interpolated to denser q- and k- grids.
    The superconducting critical temperature is obtained from the Allen-Dynes formula \cite{Allen-Dynes} by solving the Eliashberg equations on a $104\times104\times1$  q- and k- grids (see Fig. [\ref{FigureSI}] below).    
    
    The superconducting gap is evaluated by solving the anisotropic Migdal-Eliashberg equations \cite{MigdalEliashberg} in the Wannier basis on a $206\times206\times1$ grid over the imaginary frequency axis and then by performing analytic continuation to the real axis using N-point Padé approximants \cite{Vidberg1977}. In order to perform the k- and k+q- summations to solve the Migdal-Eliashberg equations on the imaginary axis, we generate random electron momenta on the Brillouin zone, and select $2688$ of them having at least one eigenvalue within $0.2$ eV of the Fermi surface. The Matsubara summation was truncated at $384$ frequencies, where convergence is reached.  We use a Morel-Anderson pseudopotential \cite{Morel-Andreson} $\mu^{*} = 0.15$ to parameterize the Coulomb repulsion in the superconducting state. The Wannier interpolation of the electron-phonon matrix elements, as well as the solution of the Migdal-Eliashberg equations have been performed within the open-source software EPI\textit{q} (Electron-Phonon Interpolation over q- and k-points) \cite{EPIQ}. 

\newpage

\begin{comment}

\end{comment}
\bibliography{references}